# Sn-Doping in LPCVD Grown (010) β-$Ga_2O_3$ Films

Saleh Ahmed Khan, Ahmed Ibreljic, Sourav Sarker, Anhar Bhuiyan*

Department of Electrical and Computer Engineering, University of Massachusetts Lowell, MA 01854, USA

*Corresponding author Email: anhar_bhuiyan@uml.edu

**Abstract**

In this work, Sn-doped (010) β-$Ga_2O_3$ homoepitaxial films were grown by low-pressure chemical vapor deposition (LPCVD), and the influence of Sn incorporation on the structural, morphological, and electrical properties was systematically investigated. Controlled room-temperature carrier concentrations ranging from $1.17 \times 10^{17}$ to $3.06 \times 10^{18}$ cm$^{-3}$ were achieved with corresponding Hall mobilities of 113-63 cm$^2$ V$^{-1}$ s$^{-1}$. The films exhibited the monoclinic β-$Ga_2O_3$ phase, near-stoichiometric composition, and well-defined step-flow morphology, with a minimum rocking-curve FWHM of 68.4 arcsec and RMS roughness of 2.63 nm. Film thicknesses from 1.66 to 11.3 μm were obtained at growth rates of 6.4-16.6 μm h$^{-1}$, demonstrating the ability of LPCVD to produce thick epitaxial layers. The sample with a room-temperature carrier concentration of $1.17 \times 10^{17}$ cm$^{-3}$ exhibited room-temperature and low-temperature Hall mobilities of 113 cm$^2$ V$^{-1}$ s$^{-1}$ and 380 cm$^2$ V$^{-1}$ s$^{-1}$ (84 K), respectively, both representing the highest reported values for LPCVD-grown Sn-doped β-$Ga_2O_3$. Transport modeling of the same sample yielded a shallow donor activation energy of 32.7 meV, a deeper donor level at 95 meV, and a low compensating acceptor concentration of $2.0 \times 10^{16}$ cm$^{-3}$, indicating efficient donor activation and a low degree of compensation. The results demonstrate that LPCVD enables controlled Sn doping while maintaining excellent structural and electrical quality, providing a viable route for the realization of thick β-$Ga_2O_3$ epitaxial drift layers.

## I. Introduction

β-$Ga_2O_3$ has emerged as one of the most promising ultrawide bandgap semiconductors for next-generation high-voltage power electronics owing to its large bandgap (4.8 eV), high critical breakdown field (6-8 MV/cm), and the availability of large-area native substrates produced by cost-effective melt-growth techniques [1-5]. These unique advantages have enabled the development of high-performance vertical power devices with breakdown voltages exceeding 10 kV, including Schottky barrier diodes, p-n heterojunction diodes, and field-effect transistors [6-20]. The realization of such devices critically depends on the availability of high-quality epitaxial drift layers with precise control over crystal quality, surface morphology, thickness, and carrier concentration. To date, β-$Ga_2O_3$ epitaxy has been demonstrated using several growth techniques, including metal-organic chemical vapor deposition (MOCVD) [21-37], molecular beam epitaxy (MBE) [38-48], halide vapor phase epitaxy (HVPE) [49-53], and low-pressure chemical vapor deposition (LPCVD) [54-64], each offering distinct advantages for material synthesis and device development.

The optimization of high-quality β-$Ga_2O_3$ epitaxial layers has predominantly focused on Si-doped films [23, 24, 26, 27, 29, 31-35], enabling low donor concentrations and electron mobilities approaching the theoretical limit. Along with Si, germanium (Ge) [37, 41, 47, 58, 63, 64] and tin (Sn) [21, 22, 39, 40, 43, 44, 61] are the primary group-IV shallow donor dopants for β-$Ga_2O_3$ and have been extensively investigated for controlled n-type doping [65]. While Si remains the most widely studied donor for epitaxial growth, Sn has emerged as a particularly attractive alternative because of its favorable substitutional incorporation on the octahedrally coordinated Ga(II) site, whereas Ge preferentially occupies the tetrahedrally coordinated Ga(I) site [65-67]. Furthermore, the ionic radius of $Sn^{4+}$ (0.69 Å) closely matches that of octahedrally coordinated $Ga^{3+}$ (0.62 Å), facilitating substitutional incorporation with minimal lattice distortion [68]. Consequently, Sn-doped β-$Ga_2O_3$ epilayers grown by MOCVD [21, 22, 69] and MBE [39, 40, 43, 44] have demonstrated controllable donor incorporation over a wide carrier concentration range. In addition, Sn is the dominant donor used in melt-grown β-$Ga_2O_3$ bulk crystals and commercially available conductive substrates,[70, 71] highlighting its technological importance for both substrates and epitaxial device structures. Although LPCVD has recently been employed to grow Sn-doped β-$Ga_2O_3$ epilayers on sapphire substrates, [72, 73] our recent demonstrations of device-quality Sn-doped β-$Ga_2O_3$ Schottky diodes employing LPCVD-grown homoepitaxial drift layers [61] highlight the potential of this growth technique. However, the effects of Sn incorporation on the structural quality, surface morphology, donor activation, and electrical transport of LPCVD-grown homoepitaxial β-$Ga_2O_3$ have not yet been systematically investigated.

In this work, we present a comprehensive investigation of Sn incorporation in LPCVD-grown (010) β-$Ga_2O_3$ homoepitaxial films over a wide carrier concentration range. By systematically controlling the Sn source loading, the relationships between Sn incorporation, crystalline quality, surface morphology, donor activation, and carrier transport are established through combined structural and temperature-dependent electrical characterization. The fundamental transport properties are further elucidated using donor activation and carrier scattering analyses, providing quantitative insight into the electrical behavior of LPCVD-grown Sn-doped β-$Ga_2O_3$. Finally, the material quality and electrical transport characteristics are benchmarked against previously reported Sn-doped β-$Ga_2O_3$ epilayers, establishing LPCVD as a viable epitaxial platform for the growth of high-quality, thick drift layers for high-voltage β-$Ga_2O_3$ power devices.

## II. Experimental Details

Sn-doped β-$Ga_2O_3$ epitaxial films were grown on Fe-doped (010) β-$Ga_2O_3$ substrates using a custom-built LPCVD system. Prior to growth, the substrates were sequentially cleaned with acetone, isopropyl alcohol (IPA), and deionized (DI) water, followed by nitrogen blow drying before loading into the reactor. The chamber temperature and growth pressure were maintained at 1000 °C and ~1.5 Torr, respectively. Ga and Sn metals, placed in separate quartz crucibles, served as the source material and dopant, respectively. Argon was used as both the purge and carrier gas, while oxygen served as the oxidizing precursor. The Sn source was positioned 2.5 cm downstream of the Ga source. The Sn donor concentration was primarily controlled by varying the Sn source loading. For selected samples, the Ga source-to-substrate distance and growth duration were also adjusted to investigate the influence of growth rate and film thickness, while the substrate temperature, chamber pressure, and gas flow rates were maintained constant. The corresponding Ga source-to-substrate distances, Sn source loadings, film thicknesses, growth rates, and room-temperature electrical properties are summarized in Table 1.

A JEOL JSM-7401F field-emission scanning electron microscope (FESEM) was used to evaluate the surface morphology. Film thicknesses were estimated from cross-sectional FESEM images of co-loaded β-$Ga_2O_3$ films grown on sapphire substrates. Surface roughness was characterized using an atomic force microscope (AFM) (Park XE-100). High-resolution X-ray diffraction (XRD) measurements were performed using a Rigaku SmartLab diffractometer with Cu Kα radiation (λ = 1.5418 Å) to evaluate the crystalline structure and quality of the epilayers. Raman spectroscopy was carried out using a Horiba LabRAM Evolution Multiline Raman spectrometer with a 532 nm excitation laser. The chemical composition and Ga/O stoichiometry of the films were analyzed by X-ray photoelectron spectroscopy (XPS) using a Thermo Scientific spectrometer equipped with a monochromated Al Kα X-ray source (photon energy = 1486.6 eV). For Hall effect measurements, Ti/Au (30/100 nm) contacts were deposited at the four corners of each sample by electron-beam evaporation, followed by rapid thermal annealing at 470 °C for 1 min in a nitrogen ambient to form ohmic contacts. The electrical transport properties were measured using an Ecopia HMS-5300 Hall effect system in the van der Pauw configuration over a temperature range of 80 to 350 K.

## III. Results and Discussion

Figure 1 presents the surface morphology of the LPCVD-grown Sn-doped (010) β-$Ga_2O_3$ homoepitaxial films (Samples 1-3) with comparable thicknesses of ~6.5 μm. The Sn source loadings of 2.17, 2.76, and 5.03 wt.% resulted in room-temperature carrier concentrations of 1.17 $\times 10^{17}$, 3.28 $\times 10^{17}$, and 3.06 $\times 10^{18}$ $cm^{-3}$, respectively. As shown in the FESEM images in Fig. 1(a), (c), and (e), all films exhibit continuous surfaces with well-defined elongated step-terrace structures aligned along the [001] direction, characteristic of step-flow homoepitaxial growth on (010) β-$Ga_2O_3$ substrates [54, 59, 74]. At the lowest Sn loading of 2.17 wt.%, the terraces appear highly uniform with well-defined step edges. Increasing the Sn loading to 2.76 wt.% produces only a slight reduction in terrace uniformity while preserving the overall step-flow morphology. At the highest Sn loading of 5.03 wt.%, corresponding to a carrier concentration of 3.06 $\times 10^{18}$ $cm^{-3}$, the terrace features become broader and more pronounced. These observations indicate that increasing Sn incorporation gradually modifies the surface morphology without disrupting the step-flow growth mode or introducing secondary surface features. The corresponding AFM images in Fig. 1(b), (d), and (f) further support this trend, showing a systematic increase in RMS roughness from 2.63 to 3.03 and 4.87 nm as the carrier concentration increases. This behavior is consistent with previous observations in LPCVD-grown Si-doped β-$Ga_2O_3$ homoepitaxial films, where increasing

Si incorporation likewise resulted in a gradual increase in surface roughness [59]. Despite this gradual increase in roughness, all films maintain relatively smooth surfaces with well-defined terraces, demonstrating that high Sn incorporation can be achieved while preserving excellent surface morphology.

To investigate the crystalline structure and crystalline quality of the LPCVD-grown Sn-doped $\beta$-$Ga_2O_3$ homoepitaxial films, high-resolution XRD was performed, as shown in Fig. 2. The $\omega$-$2\theta$ scan of the representative film with a room-temperature carrier concentration of $1.17 \times 10^{17}$ $cm^{-3}$ (Sample 1) shown in Fig. 2(a) exhibits a strong (020) $\beta$-$Ga_2O_3$ diffraction peak, confirming phase-pure $\beta$-$Ga_2O_3$ homoepitaxial growth. The crystalline quality was further evaluated from the (020) rocking curves of Samples 1-3 shown in Fig. 2(b). The full width at half maximum (FWHM) increased gradually from 68.4 arcsec for $N_D = 1.17 \times 10^{17}$ $cm^{-3}$ to 72.3 arcsec for $N_D = 3.28 \times 10^{17}$ $cm^{-3}$ and 86.7 arcsec for $N_D = 3.06 \times 10^{18}$ $cm^{-3}$, indicating a slight increase in rocking-curve broadening with increasing Sn incorporation. Nevertheless, all three samples exhibit narrow rocking-curve widths, demonstrating good crystalline quality across the investigated doping range. A similar increase in rocking-curve FWHM with increasing donor incorporation has also been reported for LPCVD-grown Si-doped $\beta$-$Ga_2O_3$ homoepitaxial films, as well as Sn-doped $\beta$-$Ga_2O_3$ films grown by MOCVD and MBE [21, 39, 59, 75]. This trend is also consistent with the gradual increase in surface roughness observed in Fig. 1.

To further evaluate the structural integrity of the Sn-doped $\beta$-$Ga_2O_3$ epilayers, room temperature Raman spectroscopy was performed. Figure 3 compares the Raman spectra of a representative LPCVD-grown Sn-doped $\beta$-$Ga_2O_3$ film with a carrier concentration of $3.14 \times 10^{17}$ $cm^{-3}$ and a thickness of 11.3 μm against the bare (010) $\beta$-$Ga_2O_3$ substrate. The Raman spectrum exhibits all characteristic Raman-active phonon modes expected for monoclinic $\beta$-$Ga_2O_3$. At the Γ-point, the optical phonons are described by the irreducible representation $\Gamma^{opt} = 10A_g + 5B_g + 4A_u + 8B_u$, where the $A_g$ and $B_g$ modes are Raman active and the $A_u$ and $B_u$ modes are infrared active [76]. Consistent with the (010) crystal orientation, the $B_g$ modes exhibit relatively weak intensities, whereas all $A_g$ modes are clearly resolved, with characteristic peaks observed at 111.4 $cm^{-1}$ ($A_g^1$), 170.2 $cm^{-1}$ ($A_g^2$), 201.2 $cm^{-1}$ ($A_g^3$), 319.9 $cm^{-1}$ ($A_g^4$), 346.8 $cm^{-1}$ ($A_g^5$), 416.1 $cm^{-1}$ ($A_g^6$), 476.7 $cm^{-1}$ ($A_g^7$), 631.1 $cm^{-1}$ ($A_g^8$), 658.7 $cm^{-1}$ ($A_g^9$), and 766.8 $cm^{-1}$ ($A_g^{10}$). The measured peak positions are in good agreement with previously reported Raman spectra of β-$Ga_2O_3$ [21, 57, 59, 62]. The close agreement between the film and substrate spectra confirms preservation of the monoclinic β-phase without detectable secondary phases or substantial changes in the Raman-active phonon modes.

Figure 4 presents the XPS characterization of a representative Sn doped β-$Ga_2O_3$ film ($N_D = 3.14 \times 10^{17}$ $cm^{-3}$) to investigate its chemical composition and bonding states. The survey spectrum shown in Fig. 4(a) confirms the presence of the constituent Ga and O elements together with weak Sn 3d and Sn 3p core-level peaks, consistent with successful Sn doping of the β-$Ga_2O_3$ films and in agreement with previous reports on Sn-doped β-$Ga_2O_3$ epilayers [36, 72, 73]. The high-resolution O 1s spectrum in Fig. 4(b) can be deconvoluted into two components centered at approximately 530.8 and 532.2 eV. The dominant lower-binding-energy peak is assigned to lattice oxygen bonded to Ga (Ga-O) in β-$Ga_2O_3$, while the higher-binding-energy component is associated with surface hydroxyl species (O-H) and adsorbed oxygen-containing surface groups. The corresponding Ga 3d spectrum shown in Fig. 4(c) is dominated by the Ga-O bonding peak centered near 20.3 eV, accompanied by a weaker higher-binding-energy component associated with surface hydroxyl-related species. Quantification of the sensitivity-factor-corrected O 1s and Ga 3d peak areas yielded oxygen and gallium atomic concentrations of 60.15% and 39.85%, respectively, corresponding to an O/Ga ratio of 1.51, in excellent agreement with the stoichiometric composition of β-$Ga_2O_3$.

Figure 5 compares the surface morphology of Samples 4 and 5 with film thicknesses of 11.3 and 1.66 μm, respectively, to evaluate the influence of film thickness on the LPCVD-grown Sn-doped β-$Ga_2O_3$ homoepitaxial films. The FESEM images in Fig. 5(a) and (c) show that both samples maintain smooth, continuous surfaces with characteristic step-terrace morphology, indicating that the step-flow growth mode is preserved over a wide thickness range. Compared with Sample 5 (1.66 μm, $N_D = 2.52 \times 10^{18}$ cm$^{-3}$), Sample 4 (11.3 μm, $N_D = 3.14 \times 10^{17}$ cm$^{-3}$) exhibits more pronounced and well-developed step-terrace features, suggesting continued terrace evolution during prolonged growth. The corresponding AFM images in Fig. 5(b) and (d) reveal low RMS roughness values of 0.94 and 1.73 nm for Samples 5 and 4, respectively. Despite the sevenfold increase in film thickness, both films retain smooth surfaces with well-defined terraces, demonstrating that thick β-$Ga_2O_3$ epilayers can be grown by LPCVD without significant degradation of the surface morphology.

Figure 6 benchmarks the room-temperature Hall mobility of the LPCVD-grown Sn-doped β-$Ga_2O_3$ homoepitaxial films against previously reported Sn-doped β-$Ga_2O_3$ epilayers grown by other growth methods such as MOCVD and MBE [21, 22, 39, 40, 43, 44]. The films developed in this work span a broad carrier concentration range from $1.17 \times 10^{17}$ to $3.06 \times 10^{18}$ cm$^{-3}$ while exhibiting room-temperature Hall mobilities between 63 and 113 cm$^2$ V$^{-1}$ s$^{-1}$. The highest Hall mobility of 113 cm$^2$ V$^{-1}$ s$^{-1}$ at $N_D = 1.17 \times 10^{17}$ cm$^{-3}$, represents the highest reported value for LPCVD-grown Sn-doped β-$Ga_2O_3$. As expected, the Hall mobility decreases with increasing carrier concentration because of enhanced ionized impurity scattering associated with increased donor incorporation. Nevertheless, the measured mobilities remain comparable to those reported for Sn-doped β-$Ga_2O_3$ epilayers grown by MOCVD and MBE over a similar carrier concentration range. More importantly, while previously reported Sn-doped $\beta$-$Ga_2O_3$ epilayers were typically grown at rates of ~0.12-0.7 μm $h^{-1}$ and resulted in film thicknesses below ~1.4 μm, the LPCVD process developed in this work achieved growth rates of approximately 6.4-16.6 μm $h^{-1}$, enabling controllable film thicknesses ranging from 1.6 to 11.3 μm without compromising electrical transport properties. These results demonstrate that LPCVD uniquely combines competitive electron mobility with substantially higher growth rates, making it a promising approach for the scalable growth of thick, device-quality $\beta$-$Ga_2O_3$ epitaxial drift layers for high-voltage vertical power devices.

To further investigate the electrical transport properties and donor activation behavior of the LPCVD-grown Sn-doped $\beta$-$Ga_2O_3$ films, temperature-dependent Hall measurements were performed on Samples 1 and 2, with room-temperature carrier concentrations of $1.17 \times 10^{17}$ and $3.28 \times 10^{17}$ $cm^{-3}$, respectively. The temperature-dependent carrier concentration and Hall mobility were analyzed using a two-donor charge neutrality model (Eq. (1)) together with temperature-dependent mobility modeling based on Matthiessen's rule (Eq. (2)) to determine the dominant carrier scattering mechanisms, donor activation energies, and compensating acceptor concentrations [77-79]. The material parameters and physical constants used in the mobility fitting are summarized in Table 2, while the extracted donor concentrations, activation energies, and compensating acceptor concentrations are listed in Table 3.

$$n + N_A = \frac{N_{D1}}{1 + 2e^{-\frac{(E_{D1}-E_F)}{k_B T}}} + \frac{N_{D2}}{1 + 2e^{-\frac{(E_{D2}-E_F)}{k_B T}}} \quad (1)$$

$$\mu_{total}^{-1} = \mu_{II}^{-1} + \mu_{NI}^{-1} + \mu_{POP}^{-1} + \mu_{ADP}^{-1} \quad (2)$$

In Eq. (1), n is the free electron concentration, $N_{D1}$ and $N_{D2}$ are the concentrations of the shallow and deep donor states with activation energies $E_{D1}$ and $E_{D2}$, respectively, $N_A$ is the compensating acceptor concentration, $E_F$ is the Fermi level, $k_B$ is the Boltzmann constant, and T is the absolute temperature. In Eq. (2), $\mu_{\text{total}}$ is the total Hall mobility, while $\mu_{II}$, $\mu_{NI}$, $\mu_{POP}$, and $\mu_{ADP}$ represent the mobility contributions limited by ionized impurity, neutral impurity, polar optical phonon, and acoustic deformation potential scattering, respectively.

Figure 7(a) shows the measured Hall mobility together with the calculated scattering contributions for Samples 1 and 2. The calculated model provides good agreement with the measured temperature dependence over the investigated range, indicating that the carrier transport is well described by the combined scattering mechanisms considered in the model. At room temperature, Hall mobilities of 113 and 96 $cm^2 V^{-1} s^{-1}$ were measured for Samples 1 and 2, respectively. As the temperature decreases, the Hall mobility initially increases because of the progressive suppression of polar optical phonon scattering, reaching peak values of 380 $cm^2 V^{-1} s^{-1}$ at 84 K for Sample 1 and 267 $cm^2 V^{-1} s^{-1}$ at 100 K for Sample 2. Upon further cooling, the Hall mobility decreases as ionized impurity and neutral impurity scattering become increasingly dominant. The extracted peak Hall mobility of 380 $cm^2 V^{-1} s^{-1}$ for Sample 1 represents the highest low-temperature Hall mobility reported for intentionally Sn-doped β-$Ga_2O_3$ epitaxial films grown by LPCVD. The measured temperature dependence and extracted scattering behavior are in good agreement with previous reports on Sn-doped β-$Ga_2O_3$ epilayers grown by MOCVD and MBE [21, 39, 44]. Figure 7(b) shows the corresponding carrier concentration as a function of 1000/T together with the fitted two-donor charge neutrality model. Both samples exhibit the characteristic donor freeze-out behavior, where the free electron concentration gradually decreases with decreasing temperature because the available thermal energy becomes insufficient to fully ionize the donor states. The excellent agreement between the experimental data and the fitted curves demonstrates that the two-donor model accurately captures the donor activation characteristics of the LPCVD-grown films. The extracted fitting parameters are summarized in Table 3. The shallow donor activation energies of 32.7 meV for Sample 1 and 26.7 meV for Sample 2 indicate efficient donor activation in the present LPCVD-grown films. The second donor activation energies of 95 and 80 meV fall within the range of deeper donor states previously reported for Sn-doped β-$Ga_2O_3$ and contribute primarily to the low-temperature carrier transport [21, 39, 44]. Furthermore, the extracted compensating acceptor concentrations of $2.0 \times 10^{16}$ and $3.5 \times 10^{16}$ $cm^{-3}$ are substantially lower than the corresponding donor concentrations, indicating a low degree of compensation. The combination of low shallow donor activation energies, low compensation, and the excellent agreement between the measured data and the transport model demonstrates the high electrical quality of the LPCVD-grown Sn-doped β-$Ga_2O_3$ homoepitaxial films.

## IV. Conclusion

In summary, high-quality Sn-doped (010) β-$Ga_2O_3$ homoepitaxial films were successfully grown by low-pressure chemical vapor deposition (LPCVD), and the effects of Sn incorporation and film thickness on the structural, morphological, and electrical properties were systematically investigated. Controlled n-type doping was achieved over a wide carrier concentration range while maintaining excellent crystalline quality, smooth step-flow surface morphology, and near-stoichiometric chemical composition, demonstrating the capability of LPCVD for the growth of high-quality Sn-doped β-$Ga_2O_3$ epilayers. Temperature-dependent Hall analysis revealed efficient donor activation, low compensation, and transport characteristics that are consistent with the dominant carrier scattering mechanisms expected in Sn-doped β-$Ga_2O_3$. The excellent agreement between the experimental measurements and transport modeling further confirms the high electrical quality of the LPCVD-grown films and provides insight into the donor activation behavior and carrier transport mechanisms. Compared with conventional epitaxial growth techniques, LPCVD offers the unique advantage of simultaneously achieving competitive electrical transport properties and substantially higher growth rates, enabling the growth of thick, device-quality β-$Ga_2O_3$ epitaxial layers without sacrificing material quality. These results establish LPCVD as a versatile and scalable epitaxial growth technique for controllable Sn doping and thick β-$Ga_2O_3$ drift-layer growth, providing a promising platform for the development of next-generation β-$Ga_2O_3$ high-power electronic devices.

**Acknowledgement**

The authors acknowledge support from the National Science Foundation (NSF) under Award Nos. ECCS-2532898 and ECCS-2501623.

**Data Availability**

The data that support the findings of this study are available from the corresponding author upon reasonable request.

**Conflict of Interest**

The authors have no conflicts to disclose.

**Table 1**

Growth parameters and electrical properties of LPCVD-grown Sn doped β-$Ga_2O_3$ films. Growths were performed at 1000 °C, with the Ga and Sn crucibles and the substrate maintained within the same thermal zone.

| Sample No. | Ga source to substrate distance (cm) | Sn/(Sn+Ga) (wt%) | Film Thickness (μm) | Growth Rate (μm/hr) | Carrier Concentration ($cm^{-3}$) | Hall Mobility ($cm^2$/V.s) |
|---|---|---|---|---|---|---|
| 1 | 8.5 | 2.17% | 6.37 | 6.37 | $1.17\times10^{17}$ | 113 |
| 2 | 8.5 | 2.76% | 6.58 | 6.58 | $3.28\times10^{17}$ | 96 |
| 3 | 8.5 | 5.03% | 6.51 | 6.51 | $3.06\times10^{18}$ | 63 |
| 4 | 6.5 | 2.69% | 11.3 | 11.3 | $3.14\times10^{17}$ | 99 |
| 5 | 6.5 | 4.92% | 1.66 | 16.6 | $2.52\times10^{18}$ | 74 |

**Table 2**

Material parameters and physical constants used in the temperature-dependent Hall mobility fitting.

| Calculation Parameter | Symbol | Value |
|---|---|---|
| Phonon Energy (meV) | $\hbar\omega_o$ | 47 (Fitted) |
| Dielectric Constant | $\varepsilon_s$ | 10.2 |
| High frequency dieletric constant | $\varepsilon_{s,\infty}$ | 3.6 |
| Electron effective mass ($m_o$) | $m^*$ | $0.313m_o$ |
| Acoustic deformation potential (eV) | $E_{ADP}$ | 6.9 |
| Mass density (kg $m^{-3}$) | $\rho$ | $5.88 \times 10^3$ |
| Sound velocity (m $s^{-1}$) | $v_s$ | $6.8 \times 10^3$ |

**Table 3**

Room-temperature electrical properties and extracted fitting parameters, including donor concentrations, donor activation energies, and compensating acceptor concentrations, for the Sn doped β-$Ga_2O_3$ films.

| Room Temperature Carrier Concentration ($cm^{-3}$) | Room Temperature Hall Mobility ($cm^2/V.s$) | $N_{D1}$ ($cm^{-3}$) | $E_{D1}$ (meV) | $N_{D2}$ ($cm^{-3}$) | $E_{D2}$ (meV) | $N_A$ ($cm^{-3}$) |
|---|---|---|---|---|---|---|
| $1.17\times10^{17}$ | 113 | $1.17\times10^{17}$ | 32.7 | $1.31\times10^{17}$ | 95 | $2.0\times10^{16}$ |
| $3.28\times10^{17}$ | 96 | $3.28\times10^{17}$ | 26.7 | $4.35\times10^{17}$ | 80 | $3.5\times10^{16}$ |

Figure 1

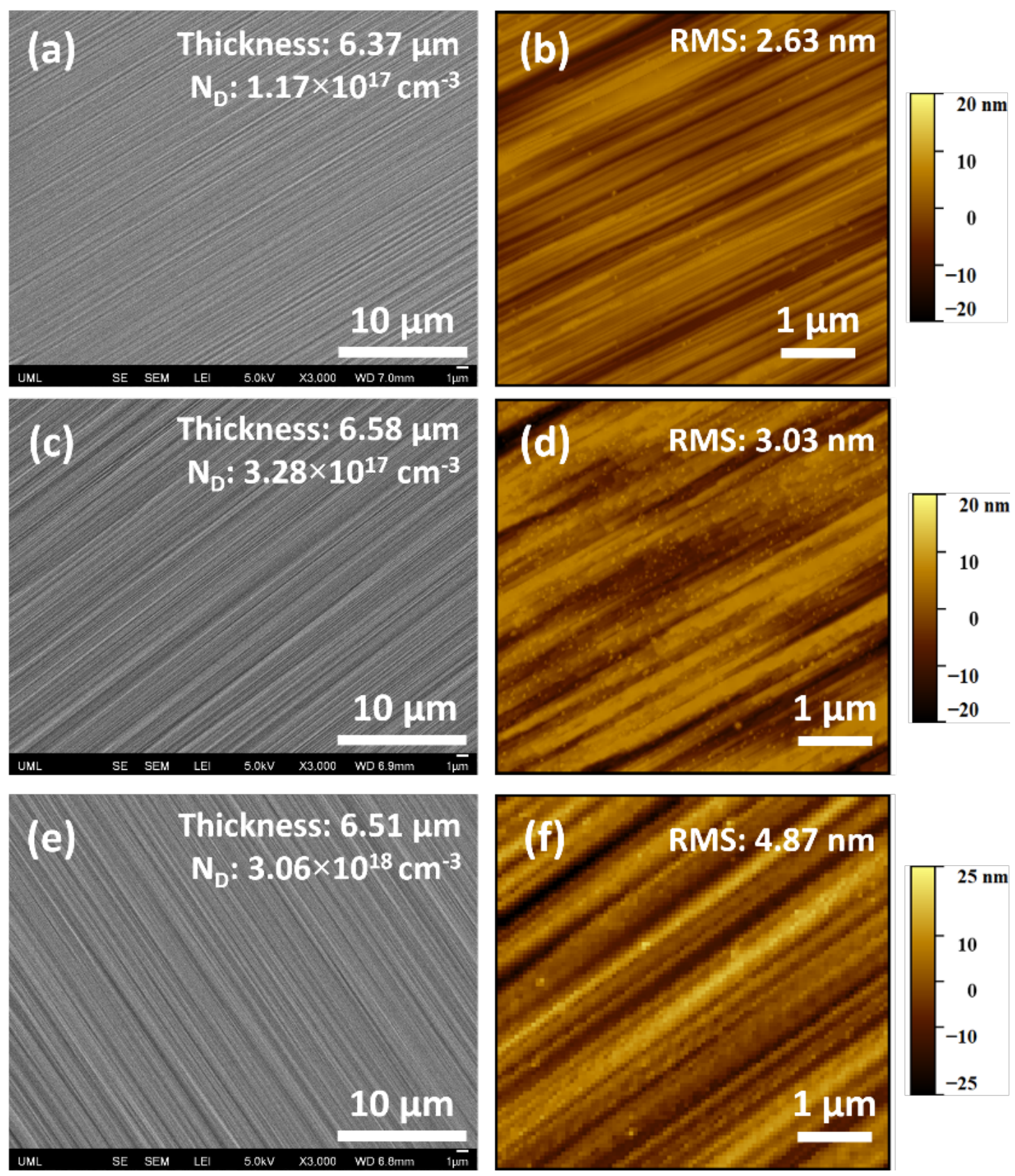

**Figure 1.** Surface morphology of LPCVD-grown Sn doped (010) β-$Ga_2O_3$ homoepitaxial films. (a), (c), and (e) FESEM images of films with room-temperature carrier concentrations of 1.17 × $10^{17}$, 3.28 × $10^{17}$, and 3.06 × $10^{18}$ $cm^{-3}$, respectively. (b), (d), and (f) corresponding AFM images acquired over a 5 µm × 5 µm scan area.

**Figure 2**

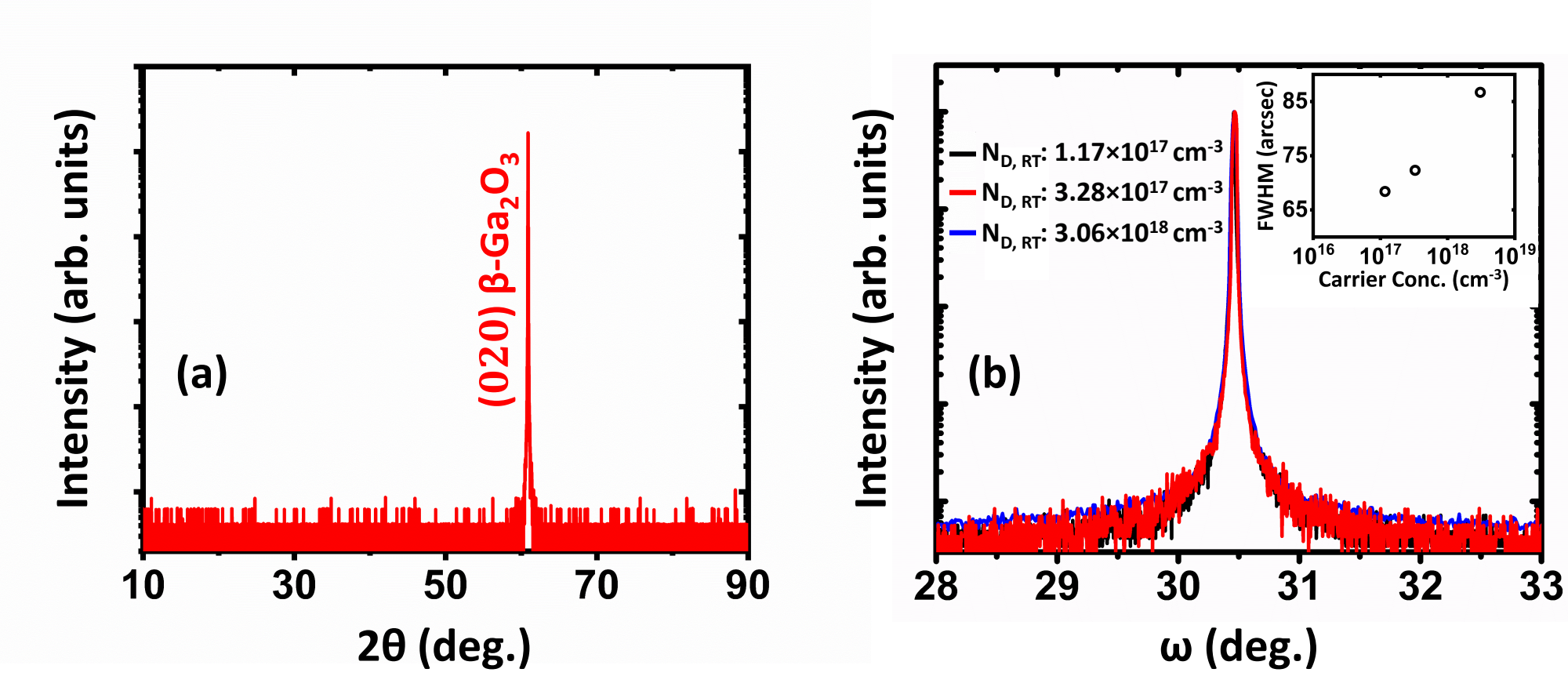

**Figure 2.** (a) High-resolution XRD ω-2θ scan of the LPCVD-grown Sn doped (010) β-$Ga_2O_3$ homoepitaxial film ($N_D = 1.17 \times 10^{17}$ $cm^{-3}$, thickness = 6.37 μm), showing a sharp (020) β-$Ga_2O_3$ diffraction peak. (b) Comparison of the (020) XRD rocking curves for Sn doped β-$Ga_2O_3$ homoepitaxial films with three different carrier concentrations. The inset shows the corresponding full width at half maximum (FWHM) of the (020) rocking curve, exhibiting a gradual increase with increasing carrier concentration.

**Figure 3**

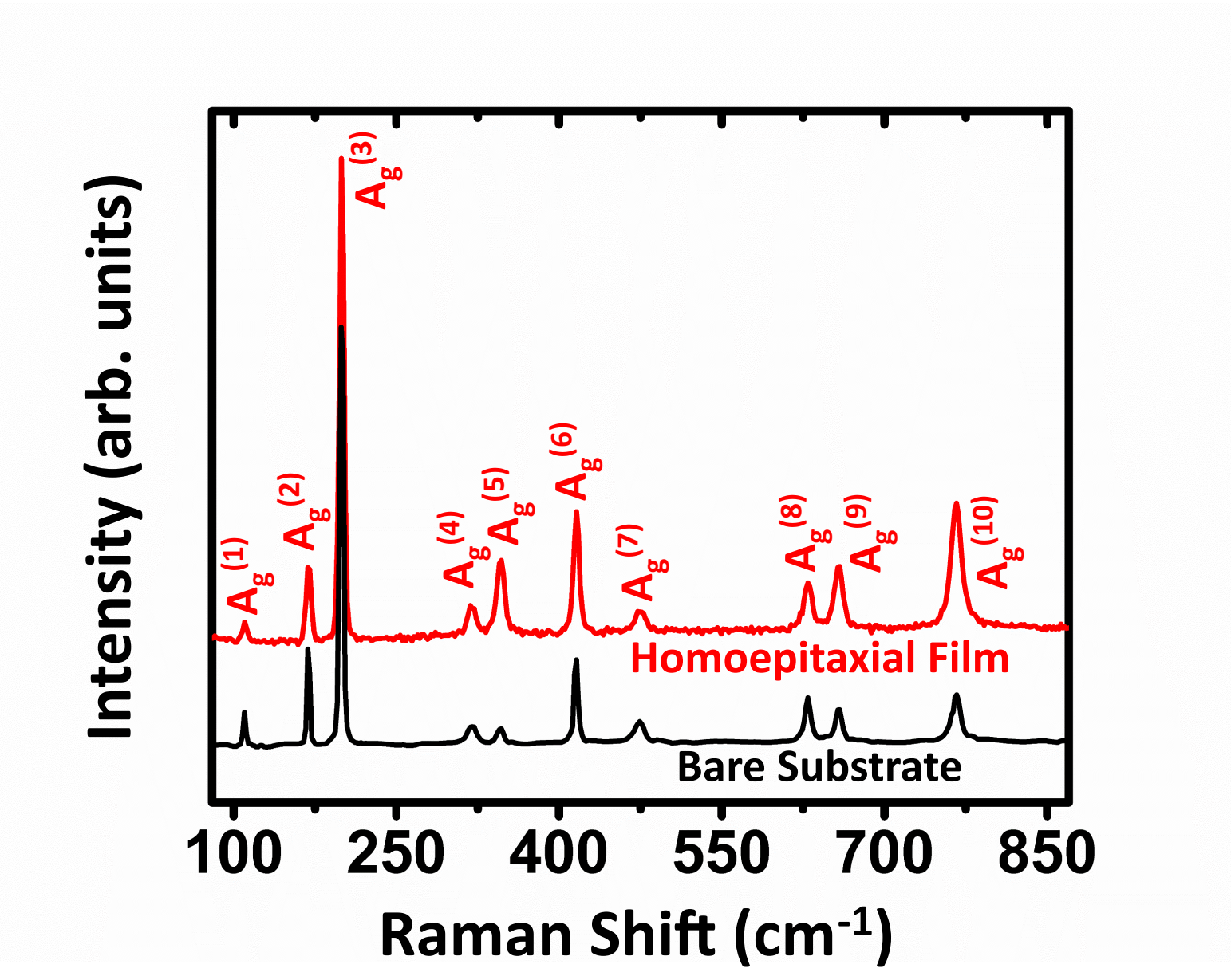

**Figure 3.** Room-temperature Raman spectra of the Sn-doped β-$Ga_2O_3$ homoepitaxial film and the bare (010) β-$Ga_2O_3$ substrate.

**Figure 4**

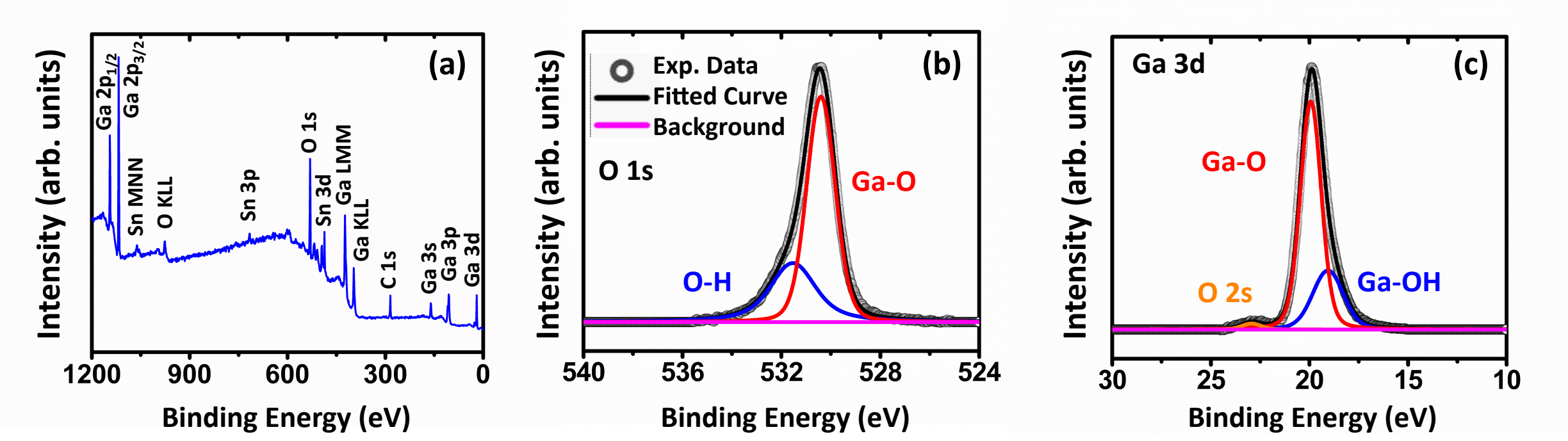

**Figure 4.** XPS characterization of the Sn-doped β-$Ga_2O_3$ homoepitaxial film. (a) Survey spectrum showing Ga, O, and Sn core-level peaks. High-resolution (b) O 1s spectrum and (c) Ga 3d spectrum.

**Figure 5**

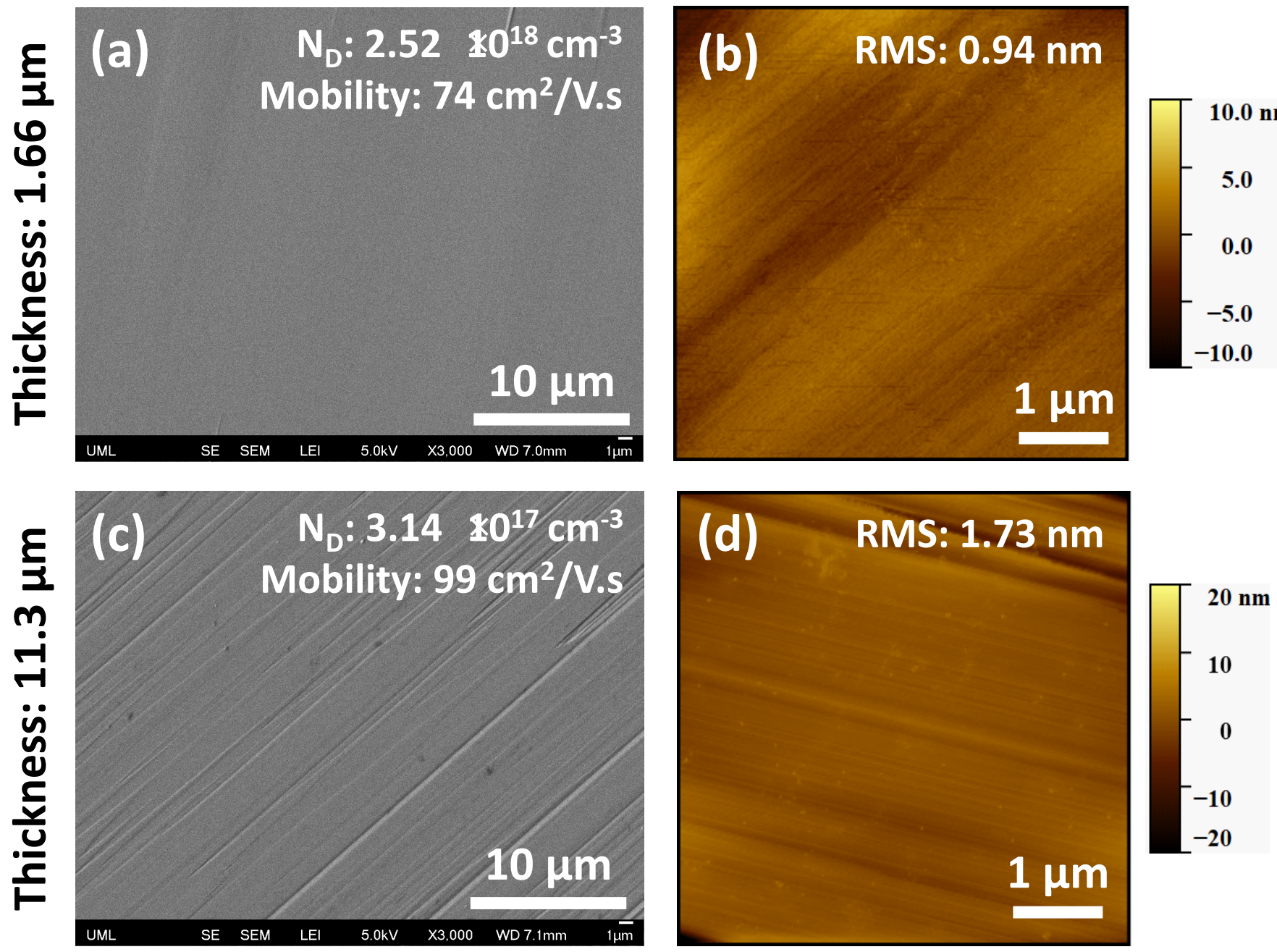

**Figure 5.** Surface morphology of LPCVD-grown Sn-doped β-$Ga_2O_3$ homoepitaxial films with different thicknesses. (a), (c) FESEM images and (b), (d) corresponding AFM images of the 1.66 and 11.3 μm thick films, respectively.

**Figure 6**

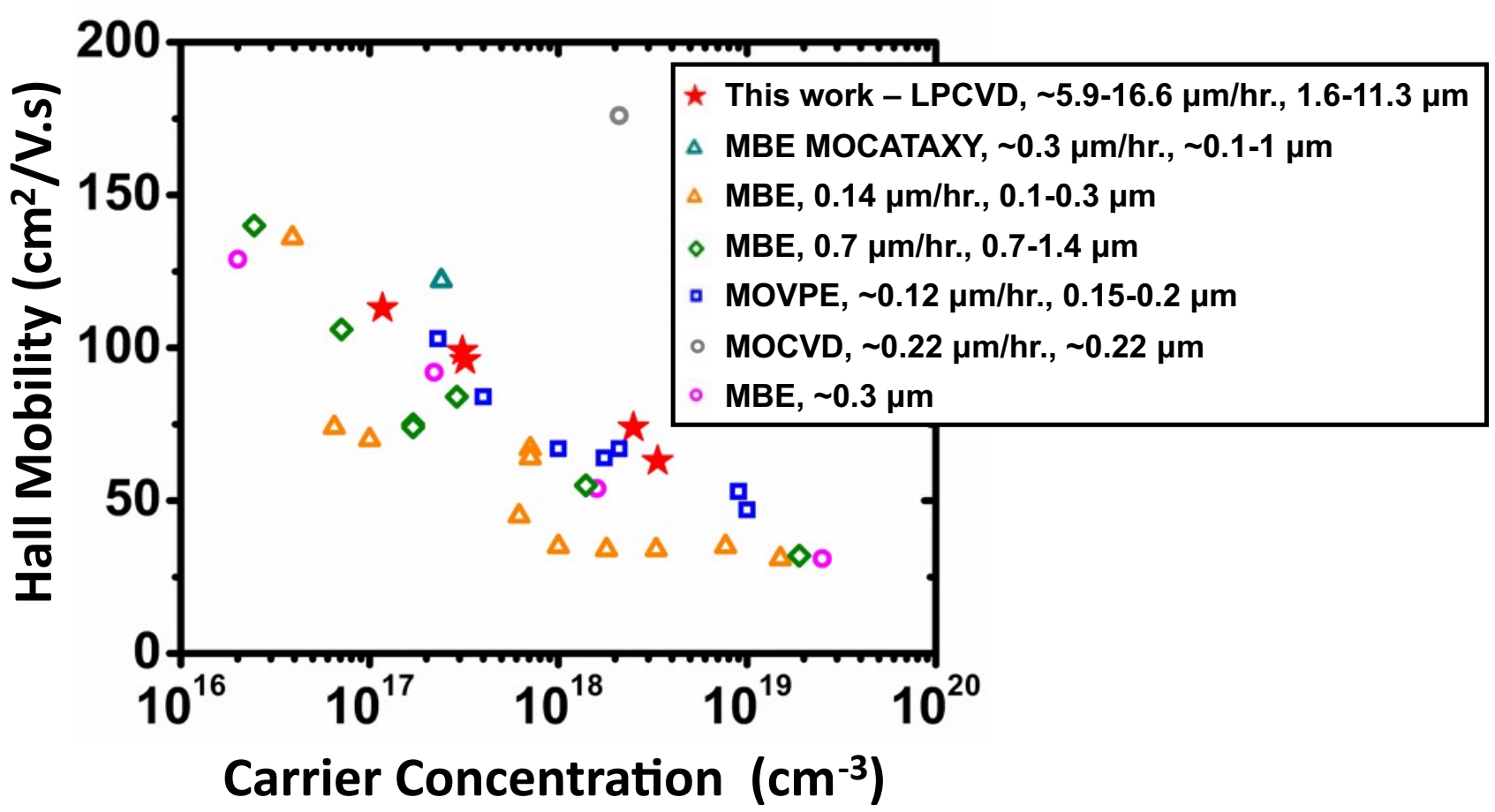

**Figure 6.** Room-temperature Hall mobility as a function of carrier concentration for LPCVD-grown Sn-doped (010) β-$Ga_2O_3$ homoepitaxial films compared with previously reported Sn-doped β-$Ga_2O_3$ epilayers grown by MOCVD and MBE. [21,22,39,40,43,44].

**Figure 7**

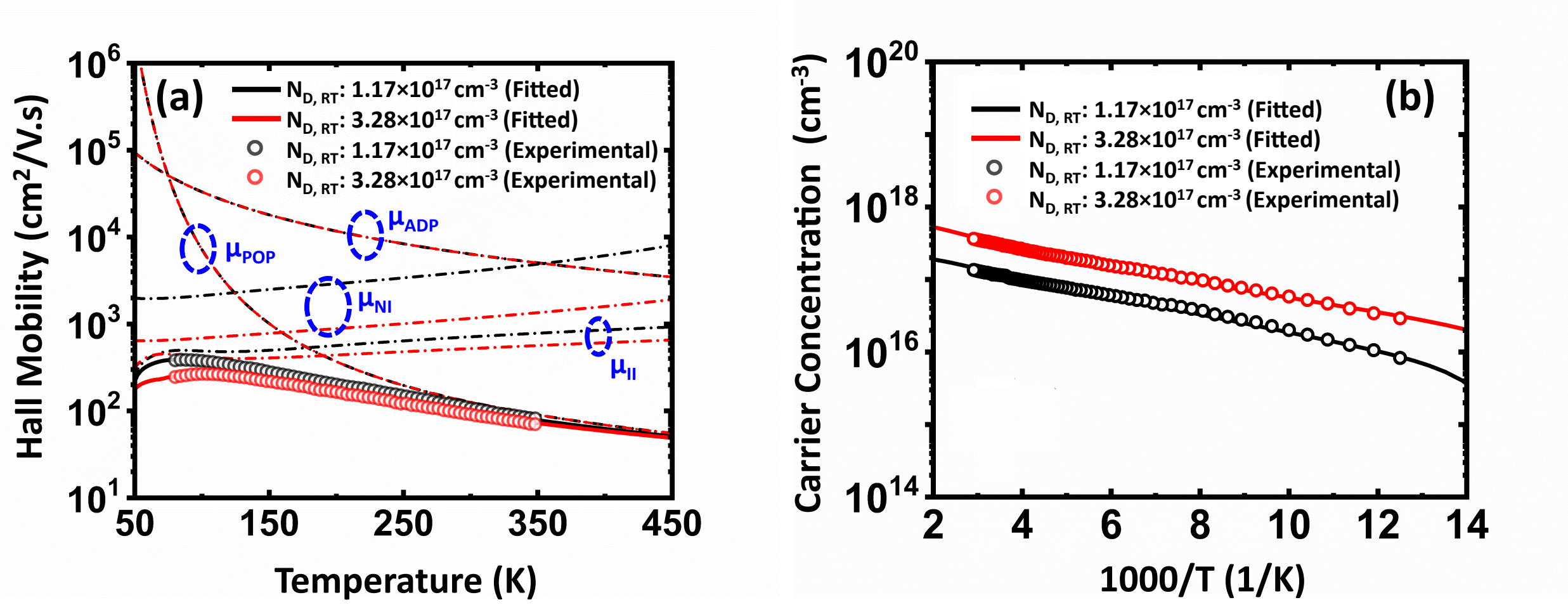

**Figure 7.** Temperature-dependent Hall transport characteristics of LPCVD-grown Sn-doped β-$Ga_2O_3$ homoepitaxial films. (a) Hall mobility with fitted scattering contributions and (b) carrier concentration with two-donor model fits for Samples 1 and 2.